\documentclass[
   pre,aps,
   superscriptaddress,
   eqsecnum,
   showpacs,
   preprintnumbers,
   byrevtex]{revtex4-1}[10pt]

\usepackage{color}
\usepackage{morefloats}
\usepackage{graphicx}
\usepackage{subfigure}
\begin{document}
\newcommand{\vphi}{\varphi}
\newcommand{\bq}{\begin{equation}}
\newcommand{\be}{\begin{equation}}
\newcommand{\ba}{\begin{eqnarray}}
\newcommand{\eq}{\end{equation}}
\newcommand{\ee}{\end{equation}}
\newcommand{\ea}{\end{eqnarray}}
\newcommand{\tchi} {{\tilde \chi}}
\newcommand{\tA} {{\tilde A}}
\newcommand{\sech} { {\rm sech}}
\newcommand{\pstar}{\mbox{$\psi^{\ast}$}}
\newcommand {\bPsi}{{\bar \Psi}}
\newcommand {\bpsi}{{\bar \psi}}
\newcommand{\sn} {\text{sn}}
\newcommand{\cn} {\text{cn}}
\newcommand{\dn} {\text{dn}}
\newcommand{\s} {\textbf{s}}
\title{Solitary wave solutions of the  2+1 and 3+1  dimensional nonlinear Dirac equation
constrained to planar and space curves}
\author{Fred Cooper}
\email{cooper@santafe.edu}
\affiliation{Santa Fe Institute, Santa Fe, NM 87501, USA}
\affiliation{Theoretical Division and Center for Nonlinear Studies, 
Los Alamos National Laboratory, Los Alamos, New Mexico 87545, USA}

\author{Avinash Khare} 
\email{khare@physics.unipune.ac.in}
\affiliation{ Physics Department, 
   Savitribai Phule Pune University, 
   Pune 411007, India}

\author{Avadh Saxena}
\email{avadh@lanl.gov}
\affiliation{Theoretical Division and Center for Nonlinear Studies, 
Los Alamos National Laboratory, Los Alamos, New Mexico 87545, USA}
\vspace{10pt}

\begin{abstract}
We study the effect of curvature and torsion on the solitons of the nonlinear Dirac equation considered on planar and space curves.  Since the spin connection is zero for the 
curves considered here, the arc variable provides a natural setting to understand the role of curvature and then we can obtain the transformation for the 1+1 dimensional Dirac 
equation directly from the metric.  Depending on the curvature, the soliton profile either narrows or expands.  Our results may be applicable to yet-to-be-synthesized curved 
quasi-one dimensional Bose condensates. 
\end{abstract} 

\date{\today}
\maketitle

\section{Introduction}
Recently  there has been renewed interest in the nonlinear Dirac equation (NLDE) because it arises in Bose condensation in honeycomb optical lattices  where confining potentials allow for quasi-one dimensional (Q1D) confinement \cite{Haddad}.  Here we show that one can find exact solitary wave solutions for the Dirac equation confined to various space curves by 
using either the arc variable representation or the vierbein formalism of Weyl \cite{Weyl}.  The latter is elucidated in the Appendix. We find that since the curves are only in the spatial part of the metric, one can transform the NLDE on the curved surface to another flat space NLDE by a coordinate transformation. This allows us to obtain the solutions of the Dirac equation analytically for the conic surfaces such as a hyperbola or parabola as well as helical (space) curves in 3+1 dimensions in terms of the spatial  arc length parameter $ \s  $.  One can furthermore analytically obtain the connection between the parametric description of the curve and the spatial  arc length variable $\s$.

%Include (i) more relevance of NLD on curves and (ii) comparison with NLS on curves. 
To our knowledge there are no studies of NLD solitons in curved geometries either experimentally or theoretically.  However, one could envision constructing curved QID Bose condensates, which serves as one of our motivations.  In contrast, solitons and breathers have been studied on curves in the context of the nonlinear Schr\"odinger (NLS) equation \cite{Gaididei,Tsironis} with an interesting interplay of curvature and nonlinearity through the soliton/breather solutions. The study of nonlinear waves on curves and closed surfaces is important in its own right \cite{ludu}. 

This paper is structured as follows.  In the next section we study the NLD equation on conic sections such as a hyperbola, parabola and ellipse.  In section III we show how the normalized charge density of a soliton is modified due to the curvature.  In section IV we consider Jacobi elliptic and other parameterizations of the various curves.  Section V deals with space curves (i.e. with finite torsion) such as a helix, an elliptic helix and the helix of the hyperboloid of revolution. Finally in section VI we provide our main conclusions. 

\section{ Nonlinear Dirac equation constrained to  space curves in $2+1$ dimensions}
 
First let us consider  the solutions of the Nonlinear Dirac Equation (NLD equation) in $2+1$ dimensions  (often called the Gross-Neveu model \cite{ref:GN}) when it is confined to a conic section which is a hyperbola, ellipse or parabola.  The NLD equation including a mass term is given by:
 
 \bq
 - i  (\gamma^\mu \partial_\mu + m)   \Psi - g^2( \bar \Psi \Psi) \Psi = 0 \,,
\eq
where $\gamma^\mu$ are Dirac matrices. It will be convenient in studying the transformation properties of this equation to introduce the quantity  $\sigma$
 \bq
 \sigma \equiv   -i  {g^2}   \bar \Psi \Psi  ,  
 \eq
 so that we can rewrite the NLD equation in the suggestive form often used in the large-N type expansions of the quantum version of this theory
  \bq  \label{nlde}
 - i  [ \gamma^\mu \partial_\mu +( m+ \sigma)]   \Psi =0. 
\eq
The fact that $\sigma$ transforms as a scalar is what will be important in what follows.

\subsection{Hyperbola} 
First let us consider the hyperbola defined by $y^2-x^2 = a^2$.  Here $\eta$ is an azimuthal angle. 
The parametrization  
\bq
y =a \sec  u , ~~~ x   =  a \tan u, 
\eq
with  $- \pi/2 < u < \pi/2$,  fulfills this condition. 
We have that 
\bq
ds^2 = dx^2 +dy^2 = \left[a^2 \sec ^2(u) \left(\tan ^2(u)+\sec ^2(u)\right) \right] du^2.
\eq

% \bq
% x= a \sinh \eta ~~ {\rm and }~~ y=  a \cosh \eta
%  \eq
% so that  the coordinates are  constrained to the hyperbola $y^2-x^2 = a^2$.

If we introduce a new coordinate via $d \s =  d s = a \sec (u) \sqrt{\tan ^2(u)+\sec ^2(u)}  du 
$, which is the arc length along the curve, we can solve the usual Dirac equation in terms of $\s$
%for arbitrary $\kappa$ 
and then use the relationship:
\bq
\s[a,u] =a (\sec (u) \sqrt{\tan ^2(u)+\sec ^2(u)}-1) 
\eq
to obtain the solutions as a function of $u, t$,  The resulting  1+1 dimensional NLD equation is 
\begin{equation}
(\gamma^0 \partial_t + \gamma^1\partial_\s  +\sigma+m)\Psi[\s,t]=0.
\label{1plus1}
\end{equation}

We see that because the spin connection is zero, one could have obtained this answer by realizing that the metric 
\ba
(ds)^2= -(dt)^2 + (dx) ^2 +(dy)^2 
= -(dt)^2 + a^2 \sec ^2(u) \left(\tan ^2(u)+\sec ^2(u)\right) du^2  \,.
\ea
The 2+1 dimensional metric  can be transformed to a $1+1$ dimensional Minkowski metric by the transformation 
\bq
 d s = a \sec (u) \sqrt{\tan ^2(u)+\sec ^2(u)}  du 
\eq
  Then  we would obtain the solutions of the NLD equation as a function of $\s,t$ and then obtain the result in terms of $u $ using $\s [u]$.  The 1+1 dimensional resulting NLDE is translationally invariant under $\s \rightarrow \s- \s_0$, so there are solitary wave solutions  centered at any value of the arc length $\s_0$. 

Note that if the transformation would have mixed space and time, such as in the light cone transformation performed in studying a scale invariant initial condition on the NLDE discussed in \cite{cooper1} this simple result would not have been possible.  Then we need to resort to a more general formulation, namely that of vierbeins discussed in the 
Appendix. 

\begin{figure}
\includegraphics[width=0.4 \linewidth]{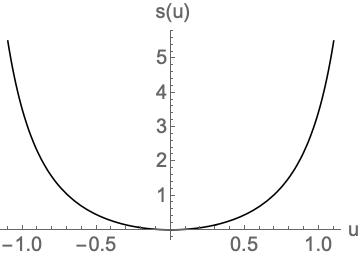} 
\caption{The curve  $ \s$ vs $ u $ for $a=1$ for the hyperbola parametrized by circular functions. } 
\label{hypt}
\end{figure}

For $a=1$ we get the curve shown in Fig. \ref{hypt}.

\subsection{NLD equation constrained to an ellipse}
\begin{figure}
\includegraphics[width=0.4 \linewidth]{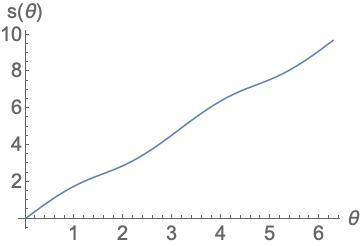} 
\caption{The arc variable curve  $ \s $ vs $ \theta $ for $a=1,~~b=2$ for the ellipse.} 
\label{ellipse}
\end{figure}

For the other conical sections the formalism is the same, just the parametric representation is changed.  
Since the spin connection is zero for these transformations, one can obtain the transformation to the 1+1 dimensional Dirac equation directly from the 
metric. 
For the ellipse centered at the coordinate origin
\bq
x= a \cos \theta \,,~~~ y= b \sin \theta \,, 
\eq
so that  
\ba
ds^2= -dt^2 + dx^2 +dy^2 
= -dt^2+(a^2 \sin^2  \theta  + b^2 \cos^2  \theta)  d\theta^2  \,. 
\ea
We see that we can reduce the metric to a $1+1$ dimensional Minkowski form by the transformation:
\bq
d \s = \sqrt{a^2 \sin^2  \theta  + b^2 \cos^2  \theta }  ~d \theta \,. 
\eq
Explicitly
\bq
\s [\theta]= b E\left(\theta \left|1-\frac{a^2}{b^2}\right.\right) \,,  
\eq
where $E(x,m)$ is the incomplete elliptic integral of the second kind with modulus $m$ \cite{AS}. When $a=1,b=2$ we get the transformation shown in Fig. \ref{ellipse}. For a circle ($a=b$) the above equation reduces to $s(\theta) = a \theta$. 

\subsection{NLD equation constrained to a parabola}
\begin{figure}
\includegraphics[width=0.4 \linewidth]{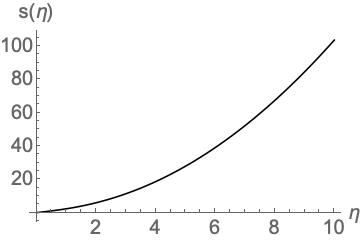} 
\caption{The arc variable curve  $ \s $ vs $ \eta $ for $a=1$ for the parabola.} 
\label{parab}
\end{figure}

For the parabola $x^2 = 4 a y$, the parametric equations are
\bq
x = 2 a \eta \,, ~~~ y= a \eta  ^2 \,.
\eq
Then 
\ba
ds^2= -dt^2 + dx^2 +dy^2 
= -dt^2+4a^2 (1+ \eta^2 )  d\eta^2 \,. 
\ea
Here 
changing variables to 
\bq 
d \s =  2a~ \int \sqrt {1+ \eta^2} ~  d \eta 
\eq
brings the metric back into the Minkowski form.
Explicitly
\bq
\s [\eta]=a  \left( \eta ~\sqrt{\eta ^2+1}+\sinh ^{-1}(\eta )\right) \,. 
\eq
When $a=1$ we obtain the curve shown in Fig. \ref{parab}. 

\section{Normalized Charge Density }
\begin{figure}
\includegraphics[width=0.4 \linewidth]{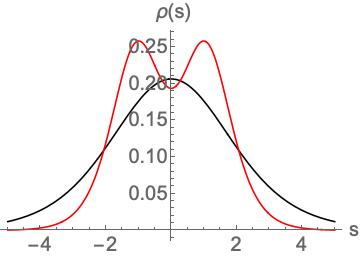} 
\caption{Normalized charge density vs. the arc length $s$  for $\omega=0.9$ (black) and $ \omega = 1/4$ (red) for NLD on a straight line.} 
\label{rhoz}
\end{figure}

What we have shown for the NLD equation in $2+1$ dimensions constrained to the three conic sections, that we can reduce the $2+1$ dimensional NLDE to the flat space $1+1$ dimensional NLDE if we use the arc length variable. 
 The static solutions of the NLDE in 1+1 dimensions  can be written in the form:
 \ba
\psi(x) &&  =  \left(  \begin{array} {cc}
      u(x) \\
      i ~v(x) \\ 
   \end{array} \right) =
R(x) \left(\begin{array}{c}\cos \theta \\ i \sin \theta \end{array}\right),
\ea
where 
\ba
u^2&&= R^2 \cos^2 \theta =  \frac{2}{g^2} \frac{(m^2-\omega^2) (m+\omega) \cosh^2 \beta x}{(m+\omega \cosh 2 \beta x)^2}, \nonumber \\
v^2&&= R^2 \sin^2 \theta =  \frac{2} {g^2} \frac{(m^2-\omega^2) (m-\omega) \sinh^2 \beta x}{(m+\omega \cosh 2 \beta x)^2}, 
\ea
as discussed in \cite{ref:numerical}  and more recently for arbitrary nonlinearity in \cite{NLD}. 
%Choosing for the parameters of the ellipse $a=1,b=2$, we have that the line
%\begin{figure}
%\includegraphics[width=0.4 \linewidth]{rhoellipse.jpg} 
%\caption{The curve  $ \s $ vs $ \theta  $ for $a=1$ for the ellipse} 
%\label{parab}
%\end{figure}
As a function of $\omega/m$ the shape of the solitary wave changes from single humped to double humped
as one lowers $\omega/m$.  As a function of the arc length $\s$ the density of the solitary wave is given by $ \rho(\s) = R^2(\s) $ 
\bq \label{A1}
\rho(\s)  = \Psi^{\dag} \Psi=\frac{2 \beta^2}{ g^2 (m+ \omega)}  \frac{1+ \alpha^2 \tanh^2 \beta \s}{(1-\alpha^2 \tanh^2 \beta \s)^2} \sech^2 \beta \s \,. 
\eq
Here (with frequency $\omega$) 
\bq
\beta = \sqrt {m^2-\omega^2} \,, ~~~ \alpha^2 = \frac{m-\omega}{m+\omega} \,, 
\eq
and the total charge is
\bq \label{A1a}
 Q=\int d\s \Psi^{\dag} \Psi
=  \frac{2 \beta}{g^2 \omega} \,. 
\eq
In Fig. \ref{rhoz}  we plot $\rho/Q[\omega]$ vs. $s$ for two values, $\omega=0.9$ which is single humped and $\omega=1/4$ which is double humped.
\\

\subsection{Hyperbola} 
\begin{figure}
\includegraphics[width=0.4 \linewidth]{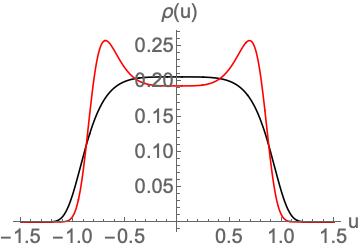} 
\caption{Normalized charge density for a soliton constrained to a hyperbola vs. $ u $ for $,\omega=0.9$ (black) and $ \omega = 1/4$ (red).} 
\label{rhohyu}
\end{figure}

We now wish to  discuss the change in shape of the solitary wave  as we go from the arc variable $\s$ to the parametric representation of the curve variable $u$  when the two dimensional motion is confined to a hyperbola.  The arc length in terms of $u $ is given by 
\bq
\s[a,u] =a (\sec (u) \sqrt{\tan ^2(u)+\sec ^2(u)}-1) \,. 
\eq
%$\s=-i a  E(i \eta |2)$ 
 In what follows we will always choose the mass  $m=1$ ,  and take $g=1$.  Later on we will use the symbol $m$ to 
describe the modulus of Jacobi elliptic functions. This is not to be confused with the Dirac mass $m$ which we set to one. 
In Fig. \ref{rhohyu}   we plot $\rho/Q[\omega]$ vs. $u$  for the same two values $\omega=0.9$ which is single humped and $\omega=1/4$ which is double humped. We see that the solitary wave shape  is quite different (i.e. contracted, see the horizontal axis) when the solitary wave is plotted vs. $u $

\subsection{Ellipse} 
\begin{figure}
\includegraphics[width=0.4 \linewidth]{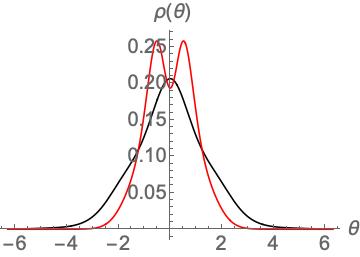} 
\caption{Normalized charge density for a soliton constrained to an ellipse with $a=1,b=2$ vs. $ \theta $ for $\omega=0.9$ (black) and $\omega = 1/4$ (red).} 
\label{rhotheta}
\end{figure}

Choosing for the parameters of the ellipse $a=1,b=2$, we have that the arc length is given by
\bq
s[\theta]=2 E\left(\theta \left|\frac{3}{4}\right.\right) \,. 
\eq
The curve for the normalized charge densities for the curve constrained to the ellipse with $b=2,a=1$  is shown in Fig. \ref{rhotheta} and is contracted.  

\subsection{Parabola} 
\begin{figure}
\includegraphics[width=0.4 \linewidth]{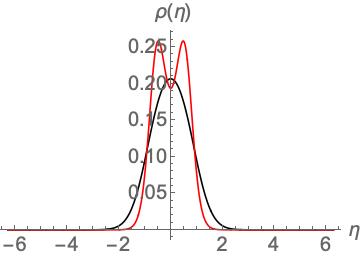} 
\caption{Normalized charge density for the  soliton constrained to a parabola  vs. $ \eta $ for $\omega=0.9$ (black) and $\omega = 1/4$ (red).} 
\label{rhoparab}
\end{figure}

Choosing for the parameters of the parabola $a=1$,  we have that the  arc length is given by
\bq
\s [\eta]= \eta  \sqrt{\eta ^2+1} +\sinh ^{-1}(\eta ) \,. 
\eq
The curve for the normalized charge densities for the curve constrained to a parabola  is shown in Fig. \ref{rhoparab}. It is even more contracted. 

%The curve for the normalized charged densities for the curve constrained to the ellipse is shown in Fig. \ref{rhotheta}. 
%\begin{figure}
%\includegraphics[width=0.4 \linewidth]{rhoellipse.jpg} 
%\caption{Normalized charge density vs. $ \theta $ for $\kappa=1,\omega=0.9$ and $\omega = 1/4$.} 
%\label{rhotheta}
%\end{figure}
%

\section{Other parameterizations of the  conic sections}
Of course there are other parameterizations of the conic sections which give different expressions for the
arc length in terms of these parameters. For example, we can parametrize the ellipse, using doubly periodic functions--namely the Jacobi elliptic functions and we can parametrize the hyperbola and the parabola using hyperbolic functions. 
\subsection{Jacobi elliptic parametrization of the ellipse} 
\begin{figure} 
\includegraphics[width=0.3 \linewidth]{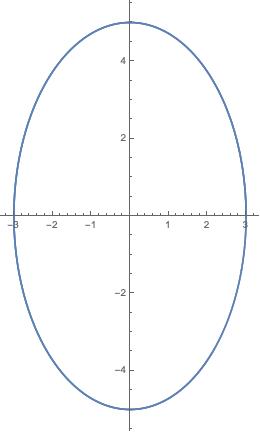} 
\caption{The Jacobi elliptic curve  $ \{x[\theta], y[\theta] \}$ for $a=3,b=5, m=1/3$  \label{Jellipse}. }
\end{figure}

\begin{figure}
\includegraphics[width=0.4 \linewidth]{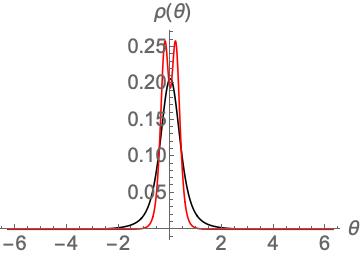} 
\caption{Normalized charge density for the  soliton constrained to an ellipse  vs. $ \theta  $ for $\omega=0.9$ (black) and $\omega = 1/4$ (red) for the Jacobi elliptic function parametrization with $b=5,a=3,m=1/3$.  It is contracted compared to Fig. 6. } 
\label{rhoee}
\end{figure}

In the simplest case we can let
\bq
x= a \cn (\theta, m) \,, ~~~ y= b \sn (\theta,m) \,, 
\eq
when the modulus $m \rightarrow 0$ the Jacobi elliptic functions $\cn(x,k) $ and $\sn(x,k)$ \cite{AS} become the circular functions $\cos(u) $ and $\sin(u)$, respectively.
%We have that the spatial part of the metric becomes
%\bq
%dx^2 + dy^2 = a^2  \text{dn}(\theta |m)^2 \eq
%so that the appropriate change of variables to a 1+1  dimensional metric  is captured in 
%\bq
%d \s = a~ \sqrt{ \text{dn}(\theta |m) }d \theta
%\eq
%or 
%\bq
%\s =a ~\text{am}(\theta |m) \theta \,, 
%   \eq
%where am($\theta$, m) is the elliptic Jacobi amplitude function. 
%%%
%\subsection{Jacobi elliptic parametrization of an ellipse}
%If instead we have 
The spatial part of the metric is
\bq
dx^2+dy^2 = \text{dn}(\theta |m)^2 \left(a^2 \text{sn}(\theta |m)^2+b^2 \text{cn}(\theta |m)^2\right) (d \theta)^2 \,, 
\eq
so that the change of variables is:
\bq
d \s = \text{dn}(\theta |m) \sqrt{a^2 \text{sn}(\theta |m)^2+b^2 \text{cn}(\theta |m)^2} d \theta \,.
\eq
Integrating we get:
\bq
 \s =  b E\left(\text{am}(\theta |m)\left|1-\frac{a^2}{b^2}\right.\right) \theta \,, 
 \eq
 which simplifies to our previous result when $b=2a$. Here ${\rm am}(\theta, m)$ is the Jacobi amplitude function \cite{AS}. The Jacobi ellipse is shown, with $a=3$, $b=5$, $m=1/3$ in Fig.~\ref{Jellipse}.  
For the densities parametrized as a function of $\theta$ of the Jacobi elliptic function we find the results shown in Fig.~\ref{rhoee}
\subsection{A different hyperbolic function parametrization of the hyperbola}
\begin{figure}
\includegraphics[width=0.4 \linewidth]{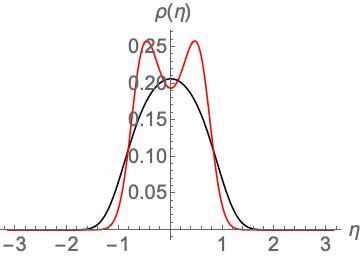} 
\caption{Normalized charge density for the  soliton constrained to a hyperbola  vs. $ \theta  $ for $\omega=0.9$ (black) and $\omega = 1/4$ (red) for the hyperbolic function parametrization with $a=2$. } 
\label{rhohyperbh}
\end{figure}

Let us consider the following parametrization of the hyperbola in terms of hyperbolic functions
\be\label{2}
x = a \sinh(\theta)\,, ~~y = a \cosh(\theta)\,. 
\ee
For this alternative parametrization of the hyperbola we find
\be
ds^2 = a^2 \cosh(2 \theta) d \theta^2 \,, 
\ee
and 
\be
s=-i a E(i u|2) \,. 
\ee
In terms of this parameterization we get the densities shown in Fig.~\ref{rhohyperbh}  when we choose $a=2$. 

%This can be generalized to an elliptic function parametrization: 
%\be\label{1}
%x = \frac{a \sqrt{m} \sn(\theta, m)}{\dn(\theta, m)}\,,~~y 
%= \frac{a}{\dn(\theta, m)}\,,
%\ee
%which again yields  $y^2 - x^2 = a^2$. Note that in 
%the limit $m =1$ it goes over to the hyperbolic mapping
%The spatial 
%part of the metric then becomes
%\be\label{3}
%dx^2 + dy^2 = \frac{m a^2 \cn^2 (\theta, m)[1+m\sn^2(\theta, m)]}
%{\dn^4(\theta, m)}\,. 
%\ee 
%Thus the appropriate change of variables to a $1+1$ dimensional metric is 
%captured in 
%\be\label{4}
%ds = \frac{a \sqrt{m} \cn(\theta, m) \sqrt{1+m\sn^2(
%\theta, m)}}{\dn^2(\theta, m)}\, d\theta\,. 
%\ee
%On making the substitution $\dn(\theta, m) = t$, we then get the integral for the arclength:
%\ba\label{5}
%s(\theta, m)&& = -a \int dt\, \frac{\sqrt{2-t^2}}{t^2 \sqrt{1-t^2}}\, \nonumber \\
%&&= -a \left[ \frac{\sqrt{\frac{t^2-1}{t^2-2}} \left(t^4-3 t^2+\sqrt{1-t^2} \sqrt{2-t^2} t
%   E\left(\left.\sin ^{-1}\left(\frac{t}{\sqrt{2}}\right)\right|2\right)+2\right)}{t
%   \left(t^2-1\right)}\right]
%\ea

\subsection{Hyperbolic function parametrization of a parabola}
\begin{figure}
\includegraphics[width=0.4 \linewidth]{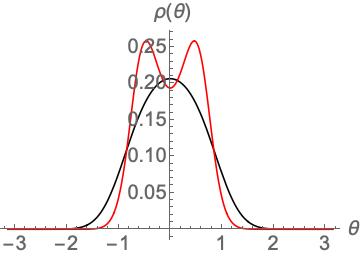} 
\caption{Normalized charge density for the  soliton constrained to a parabola  vs. $ \eta  $ for $,\omega=0.9$ (black) and $\omega = 1/4$ (red) for the hyperbolic function parametrization with $a=1$.} 
\label{rhoparabh}
\end{figure}

Instead of the parametrization given above of the parabola we could have used a parametrization in 
 terms of hyperbolic functions:
\be
x = 2a \sinh(\eta)\,,~~y = a \sinh^2(\eta)\,,
\ee
so that again  $x^2 = 4 a y$. Note that for small $\eta$ this parametrization goes over
to the parametrization used earlier. The spatial part of the metric then becomes
\be\label{7}  
dx^2+dy^2 = 4a^2 \cosh^4(\eta)~ d \eta^2 \,. 
\ee
Thus the appropriate change of variables to a $1+1$ dimensional metric is 
captured in 
\be\label{8}
ds = 2a \cosh^2(\eta) d\eta\,.
\ee
In this way, we obtain
\be\label{9}
s(\eta) = a[\eta+\sinh(2\eta)/2]\,.
\ee
  One obtains for the normalized densities in the two cases studied before the results shown in Fig.~\ref{rhoparabh}.

\section{Solitary wave solutions for the  $3+1$ dimensional NLDE confined to a space curve}
\subsection{Helix} % and Jacobi elliptic function helix 
The helix is a space curve with parametric equations:
\bq
x=r \cos \theta \,, ~~~ y= r \sin \theta \,, ~~~ z= c \theta \,. 
\eq
Here $r  
% \in [0, 2 \pi] 
$ is the radius of the helix and $2 \pi c$ gives the vertical separation of the helix's hoops, i.e. pitch.
The curvature of the helix $\kappa$ is given by
\bq
\kappa = 
\frac{r}{r^2+c^2} \,, 
\eq
which is constant and the torsion $\tau$ is given by 
\bq
\tau = \frac{c}{r^2+c^2} \,, 
\eq
which is also constant.  For this problem we find the metric is given by 
\ba
ds^2 = - d \tau^2 + dx^2 + dy^2 + dz^2  
=  - d \tau^2 + (r^2 + c^2) d \theta^2 \,. 
\ea
Therefore,  for this problem, the simple transformation  to the 1+1 Minkowski metric is again in terms of the arc length 
\bq
d \s =\sqrt {r^2 + c^2}  d \theta \,. 
\eq
So again in the $1+1$ dimensional space the transformation is given by 
\bq
d \s  = \sqrt{-g} d \theta \,.  
\eq

%If we instead have the curve defined by
%\bq 
%x=a JacobiCN[u, 1/2];~~ y= r \sin \theta;~~ z= c \theta
%\eq

\subsection{Elliptic helix}

The elliptic helix, i.e. a helix with an elliptic cross-section on projection, is defined by 

\bq
x=a \cos \theta \,, ~~~ y= b \sin \theta \,, ~~~ z= k \int \sqrt{a^2 \sin^2 \theta + b^2 \cos^2 \theta } d \theta  \,. 
\eq
Note that $dz^2$ = $k^2 (dx^2+dy^2)$ so that the 3D arc length is proportional to the 2D arc length.  The curvature is 
variable in this case. 
Performing the integral we get 
\bq
z = a E\left(\theta \left|1-\frac{b^2}{a^2}\right.\right) \,. 
\eq
The metric is given by
\ba
ds^2 = - d \tau^2 + dx^2 + dy^2 + dz^2 
=  - d \tau^2 +(1+k^2) (a^2 \cos ^2(\theta )+b^2 \sin ^2(\theta ))   d \theta^2 \,, 
\ea
so that again one can transform this to a Minkowski metric with the transformation utilizing the arc length
\bq
d \s  = \sqrt{-g} d \theta= \sqrt{1+k^2} \sqrt{a^2 \cos ^2(\theta )+b^2 \sin ^2(\theta )} ~d\theta \,. 
\eq
Thus, the explicit transformation is  (for $a>b >0$)  
\bq
\s=\sqrt{1+k^2}a E\left(\theta \left|1-\frac{b^2}{a^2}\right.\right) \,. 
\eq

\subsection{Jacobi elliptic helix}
\begin{figure} 
\includegraphics[width=0.3 \linewidth]{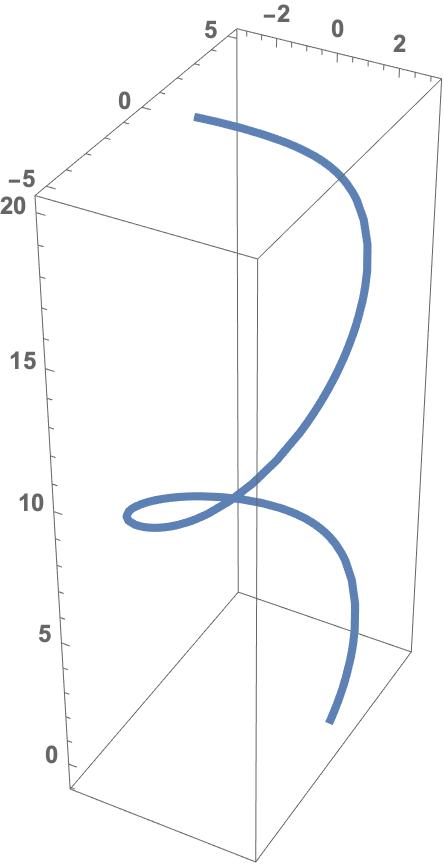} 
\caption{The Jacobi elliptic helix curve  $ \{x[\theta], y[\theta], z[\theta] \}$ for $a=3,b=5, m=1/3,k=1$.  \label{Jacobicurve} }
\end{figure}

The Jacobi elliptic helix can be parametrized as follows:
\bq
x=a \cn (\theta~|m) \,, ~~~ y= b \sn( \theta ~|m) \,, ~~~ z= k \int  d \theta ~\text{dn}(\theta |m) \sqrt{\left(a^2 \text{sn}(\theta |m)^2+b^2 \text{cn}(\theta |m)^2\right)} \,.
\eq
Performing the integral we have 
\bq
z= k b E\left(\text{am}(\theta |m)\left|1-\frac{a^2}{b^2}\right.\right) \,. 
\eq
The space curve is shown in Fig. \ref{Jacobicurve}, and 
\bq
dx^2+dy^2+dz^2 = (1+k^2) (\text{dn}(\theta |m))^2  \left(a^2 \text{sn}(\theta |m)^2+b^2 \text{cn}(\theta |m)^2\right) \,. 
\eq
So we see that we can transform this to a 1D Dirac equation by letting 
\bq
  d \s = \sqrt{1+k^2}  \text{dn}(\theta |m) \sqrt{\left(a^2 \text{sn}(\theta |m)^2+b^2 \text{cn}(\theta |m)^2\right)} d \theta \,. 
  \eq
  
Integrating, we get the arc variable 
\bq
\s = \sqrt{1+k^2} b E\left(\text{am}(\theta |m)\left|1-\frac{a^2}{b^2}\right.\right) \,. 
\eq

\subsection{Helix of the hyperboloid of revolution}
The hyperboloid of revolution is defined by 
\bq
\frac{x^2+y^2}{a^2} - \frac{z^2}{b^2}=1.
\eq

The helix of the one-sheeted hyperboloid of revolution can be parametrized as follows:
\bq \label{harc} 
x= a \cosh \eta \cos {f(\eta)} \,, ~~~y = a \cosh \eta \sin { f(\eta)} \,,~~~ z = b \sinh  \eta \,, 
\eq
where 
\bq \label{feta} 
\frac{df}{d\eta }  = \sqrt { \left(\frac{b}{a \tan \alpha} \right)^2 - \tanh^2 \eta } \,. 
\eq
The curve $f[\eta]$ depends on whether
\bq
\beta =\frac{b}{a \tan \alpha}  
\eq
is less than one or greater than or equal to one.  In the latter case $f[\eta] $ is a straight line.  In our plot, Fig.~13, we consider the cases
$\beta = 1/2, 1, 10$ to exemplify the possibilities. 
\begin{figure} 
\includegraphics[width=0.4 \linewidth]{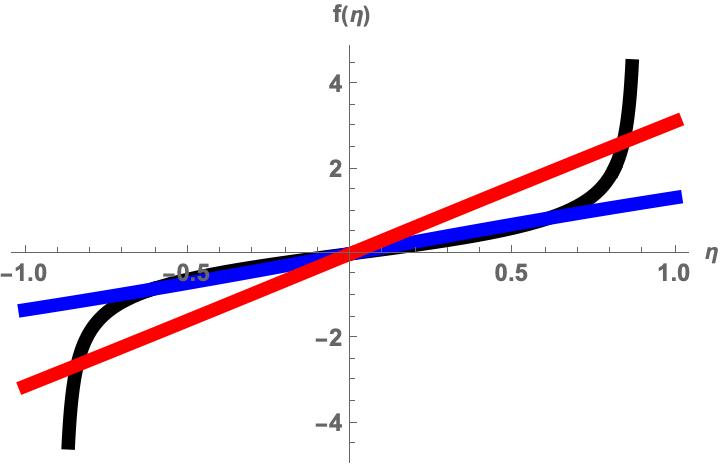} 
\caption{The curves  $ f[\eta] $ for  $\beta=1/2,1,10$,  drawn in black, blue and red. Here we have chosen $a=1$, $b=1/2$.  \label{feta} }
\end{figure}
Explicitly, 
\ba
f_{1/2}(\eta)&& = \frac{\cosh (\eta ) \left(\sqrt{6-2 \cosh (2 \eta )} \sin ^{-1}(\sinh (\eta ))-2
   \sqrt{\cosh (2 \eta )-3} \tanh ^{-1}\left(\frac{2 \sinh (\eta )}{\sqrt{\cosh (2 \eta
   )-3}}\right)\right)}{\cosh (2 \eta )-3} \,, \nonumber \\
f_1(\eta)&&=   \frac{\cosh (\eta ) \sqrt{4-2 \tanh ^2(\eta )} \left(\sinh ^{-1}\left(\frac{\sinh (\eta
   )}{\sqrt{2}}\right)+\tan ^{-1}\left(\frac{\sqrt{2} \sinh (\eta )}{\sqrt{\cosh (2 \eta
   )+3}}\right)\right)}{\sqrt{\cosh (2 \eta )+3}} \,, \nonumber \\
f_{10}(\eta)&&= \frac{\cosh (\eta ) \sqrt{20-2 \tanh ^2(\eta )} \left(3 \sinh ^{-1}\left(\frac{3 \sinh
   (\eta )}{\sqrt{10}}\right)+\tan ^{-1}\left(\frac{\sqrt{2} \sinh (\eta )}{\sqrt{9 \cosh
   (2 \eta )+11}}\right)\right)}{\sqrt{9 \cosh (2 \eta )+11}} \,. 
   \ea
From Eq.~(\ref{harc}) one can show that the differential arc length is 
\bq
 d \s =b \csc (\alpha ) \cosh (\eta) \,. 
 \eq
 Since
 \bq
 d z = b  \cosh (\eta) \,, 
 \eq
 therefore the parameter  $\alpha$ has the meaning
 \bq
 \frac{dz}{d \s} = \sin \alpha
 \eq
 and the arc length is given by 
 \bq
 \s = b \csc (\alpha ) \sinh (\eta) \,. 
 \eq
 The  helices for $\beta = 1/2,1,10$ are shown in Fig.~\ref{helices}.
 \begin{figure} 
\includegraphics[width=0.3 \linewidth]{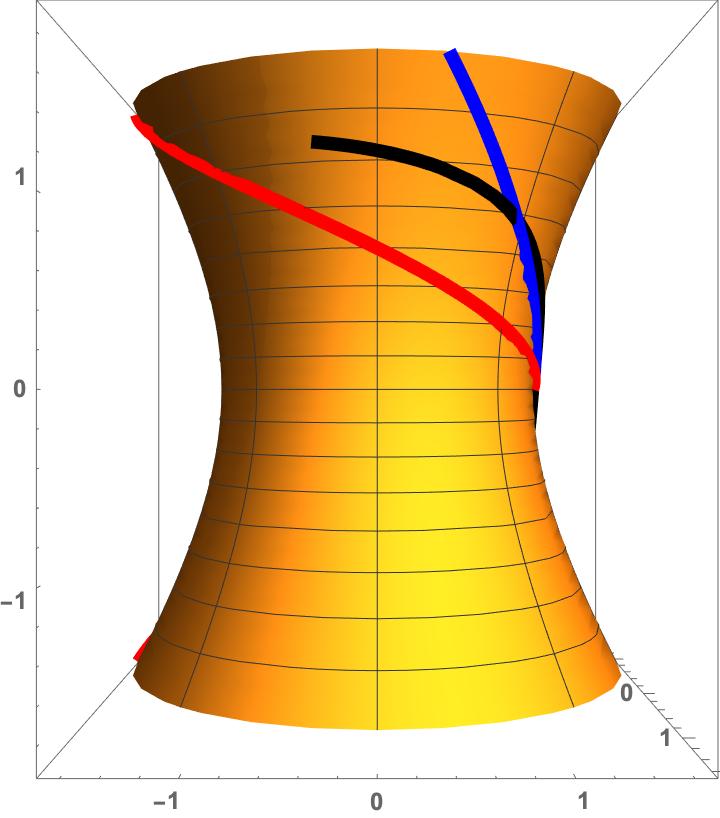} 
\caption{The helices for  $\beta=1/2,1,10$, are  drawn in black, blue and red on a one-sheeted hyperboloid of revlution. Here we have chosen $a=1$, $b=1/2$.  \label{helices} }
\end{figure}

\section{Conclusions}  
We have studied the behavior of the nonlinear Dirac equation \cite{ref:numerical, NLD} on planar and space curves.  We have shown here how the arc length variable is the relevant 
choice to study the nonlinear Dirac equation on planar and space curves.  We studied different parameterizations of various curves including those involving 
Jacobi elliptic functions \cite{AS} and studied the charge density in the presence of a soliton. We found the change in soliton shape in terms of narrowing or broadening 
of the soliton profile in the curved region.  These results illustrate an insightful interplay between solitons of the nonlinear Dirac equation and the curvature (and torsion) of a variety of  
curves. Our results are relevant to curved Q1D Bose condensates assuming they can be realized experimentally.  It would 
be instructive to (numerically) study collisions of NLD solitons on various curves to explore how curvature (and torsion) affect their interaction and collision dynamics including bounce windows. 

\section{Acknowledgment} 
A.K. is grateful to Indian National Science Academy (INSA) for  the award of INSA Senior Scientist position at Savitribai Phule Pune University. 
This work was supported in part by the U.S. Department of Energy.

 \section{ Appendix: The viebein formalism applied to a hyperbola} 
 
The vierbein introduced by Hermann Weyl \cite{Weyl} is also called a tetrad or a frame field (in general relativity). We shall use the metric convention $(-++)$ for the $2+1$ space-time which is commonly used in the
curved-space
literature. In what follows, we use Greek indices for the curvilinear
 coordinates  $\eta$  or $\theta$, and Latin
indices for the Minkowski coordinates $x,y$ and $t$. To obtain the fermion
evolution equations in the new coordinate
system it is simplest to use a coordinate covariant action such as that
used in field theory in curved spaces, even though here the actual curvature is
zero.   For the constrained system, the  $2+1$ dimensional  Minkowski line element  
%\begin{equation}
%ds^2=-d\tau^2+\tau^2 d\eta^2+ dy^2 \to g_{\mu\nu}=(-1,\tau^2,1).
%\label{eq2.3}
%\end{equation}
\ba
(ds)^2&&= -(dt)^2 + (dx) ^2 +(dy)^2 \nonumber\\ 
&&= -(dt)^2+a^2 \cosh 2 \eta  (d\eta)^2 \equiv  -(dt)^2 + (d \s)^2 \,, 
\ea
so that the original Minkowski metric 
\bq
g_{\alpha \beta }={\rm diag}\{-1,1,1\} 
\label{eq2.3}
\end{equation}
reduces to  an effective curved space $1+1$ dimensional metric tensor:
\bq
g_{\mu\nu}={\rm diag}\{-1,a^2 \cosh 2 \eta\}  
\label{eq2.3}
\end{equation}
with inverse
\bq
g^{\mu\nu}={\rm diag}\{-1, \frac{1}{a^2 \cosh 2 \eta}\}.
\label{eq2.3a}
\end{equation}

Vierbeins transform the curved 1+1 dimensional  space to a locally Minkowski  1+1 dimensional space via. 
$
g_{\mu\nu}=V^a_{\mu} V^b_{\nu} \eta_{ab},
$
where $\eta_{ab}=(-1,1)$ is the flat Minkowski metric.
A convenient choice for the vierbein is
\bq
   V_{\mu}^{a}  = {\rm diag}\{1,a \sqrt {\cosh 2 \eta} \} ~,
\end{equation}
so that
\begin{equation}
V^\mu_a= {\rm diag}\left\{1, \frac {1}{a \sqrt {\cosh 2 \eta}}\right\}\,.
\label{boost_vierbein}
\end{equation}

The determinant of the metric tensor is given by
\begin{equation}
\det V = \sqrt{-g} = a \sqrt {\cosh 2 \eta} \,.
\label{boost_detV}
\eq
The action for our model in general curvilinear coordinates is
\begin{equation}
S[\Psi,\sigma]=\int d^3x~\det V\bigg({-i\over 2}\bar{\Psi}\tilde\gamma^{\mu} \nabla_{\mu}\Psi
+{i\over 2}(\nabla_{\mu}^{\dagger}\bar{\Psi})\tilde\gamma^{\mu}\Psi- i(m \bar{\Psi}\Psi) 
- \frac{g^2}{2} ( \bar{\Psi}\Psi)^2 \bigg) \,. 
\label{action}
\end{equation}
The coordinate dependent gamma matrices
$\tilde{\gamma}^{\mu}$ are obtained from the usual Dirac gamma matrices
$\gamma^{a}$
via
\begin{equation}
 \tilde{\gamma}^{\mu} = \gamma^{a} V_{a}^{\mu}(x)\ .
\end{equation}
The coordinate independent Dirac matrices $\gamma^{a}$ satisfy the
usual gamma matrix algebra:
\begin{equation}
\{\gamma^{a},\gamma^{b}\} = 2 \eta^{ab}~.
\label{boost_gammaD}
\end{equation}

{}From the action Eq. (\ref{action}) we obtain the Heisenberg field
equation for the fermions,
\begin{equation} \label{nlde}
\left( \tilde{\gamma}^{\mu}\nabla_{\mu} + \sigma+m \right) \Psi=0\,,
\end{equation}
%which takes the form
%\begin{equation}
%\left[ \gamma^0 \left(\partial_\tau+\frac{1}{2\tau}\right)
%+ \frac{\gamma^3}{\tau} \partial_\eta + \sigma  \right] \Psi_i =0 ~.
%\label{boost_Dirac}
%\end{equation}
where it is to be understood: 
\begin{equation} \label{constraint}
\sigma  = - i g^2 \bar  \Psi  \Psi \,. 
\end{equation}
This can be generalized easily to the case where the nonlinear term in the NLDE is of the form \cite{NLD} 
\begin{equation}
 -  g^2  (\bar  \Psi  \Psi ) ^\kappa \Psi \,,
\end{equation}
with $\kappa$ denoting arbitrary nonlinearity. 
This is done by letting 
\bq \label{sk}
\sigma=  - i g^2 (\bar  \Psi  \Psi )^\kappa
\eq 
in the equation of motion.  This again is space and time dependent mass, but a scalar as we will show below. 

Here  $\tilde\gamma^{\mu}=\gamma^aV_a^{\mu}$, $\nabla_{\mu}=\partial_{\mu}+\Gamma_{\mu}$,
and $\Gamma_{\mu}$ is the spin connection given by \cite{Birrell_Davies,Weinberg}
\begin{equation}
\Gamma_{\mu}={1\over 2}\Sigma^{ab}V_{a\nu}(\partial_{\mu} V_b^{\nu}+\Gamma^{\nu}_{\mu\lambda}V_b^{\lambda}) \,, 
~~~\Sigma^{ab}={1\over 4}[\gamma^a,\gamma^b].
\label{eq2.10}
\end{equation}
We have
\bq
 \tilde{\gamma}^{0}= \gamma^0 ; ~~  \tilde{\gamma}^{1}= \frac{1}{a \sqrt {\cosh 2 \eta}} \gamma^1 \,. 
 \eq
 
 We will choose for convenience the following representation
for the matrices $\gamma^0$ and $ \gamma^1$ in the transformed local Minkowski 1+1 dimensional Dirac equation. 
\begin{equation}
 i \gamma^0 =   \left( \matrix{ 1  &  0 \cr
0  & -1  \cr}  \right) \,, 
\end{equation}

\[
\gamma^1 =  \left( \matrix{ 0 & 1 \cr
1 & 0 \cr} \right) \,.
\]

 We have that the scalar field $\sigma$ is unchanged in form
 \bq
 \sigma = - i g^2 ( \Psi ^\dag  \tilde{\gamma}^{0}\ \Psi ) =  - i g^2 ( \Psi ^\dag  {\gamma}^{0}\ \Psi ) \,, 
\eq
if we used Eq. (\ref{sk}) to define $\sigma$, it would again be unchanged in form.
 
The Christoffel symbols have the usual definition,
$\Gamma^{\sigma}_{\mu\lambda}= \frac{1}{2} g^{\nu \sigma} (\partial_\lambda  g_{\nu \mu}+\partial_{\mu}g_{\nu \lambda}
-\partial_\nu  g_{\mu \lambda})$.
Since the only non-zero derivative is 
\bq
\frac{\partial g_{22}}{\partial x^2} = 2 a^2 \sinh 2 \eta,
\eq
we find the only nonzero Christoffel symbol is 
\bq
\Gamma_{222} = a^2 \sinh 2 \eta
\eq
so that 
\bq
\Gamma^2_{22} = \tanh 2 \eta \,. 
\eq

We find the spin connection is zero for the following reason. 
If we consider 

\bq
F_{\mu, b} ^\nu = \partial_{\mu} V_b^{\nu}+\Gamma^{\nu}_{\mu\lambda} V_b^{\lambda} \,, 
\eq
only $F_{22}^2$  could have a  non-zero  contribution but  we find \bq
\partial_\nu V_2^2 =- \frac {\tanh 2 \eta} { a \sqrt{\cosh 2 \eta}};~~ \Gamma_2^{22} V_2^2 = +  \frac {\tanh 2 \eta} { a \sqrt{\cosh 2 \eta}} \,, 
\eq
so that   $F_{22}^2=0$ which implies that the spin connection is zero.

~From Eq.~(\ref{nlde}), and the fact that the spin connection is zero we find that the equation of motion for $\Psi$ is
\begin{equation}
(\tilde\gamma^{\mu}\partial_{\mu}+\sigma+m)\Psi=0.
\label{eq2.14}
\end{equation}
Writing this out more explicitly, we obtain
\begin{equation}
\left(\gamma^0 \partial_t +\frac{1}{a \sqrt {\cosh 2 \eta}} \gamma^1\partial_\eta  +\sigma+m\right)\Psi[\eta,t]=0.
\label{eq2.15}
\end{equation}
%If we rescale the equation by letting $\Psi=\Phi/\sqrt{\tau}$ and
%$\tau={1\over m}e^u$, we obtain,
%\begin{equation}
%(\tau\gamma^2\partial_y+\gamma^0\partial_{u}+\gamma^1\partial_{\eta}+\tilde\sigma)\Phi
%\equiv (\hat\gamma^{\mu}\partial_{\mu}+\tilde\sigma)\Phi=0,
%\label{eq2.16a}
%\end{equation}
%where $\tilde\sigma=\sigma\tau$. 

\end{document}